\documentclass{article}

\usepackage{microtype}
\usepackage{graphicx}
\usepackage{subcaption}
\usepackage{booktabs} % for professional tables
\usepackage{enumitem}
\usepackage{multirow}
\usepackage{tabularx}
\usepackage{array}
\usepackage{multirow}
\usepackage{makecell}
\usepackage{pifont}\usepackage[hidelinks]{hyperref} % 或者不加 hidelinks

\newcommand{\cmark}{\ding{51}}

\newcolumntype{C}[1]{>{\centering\arraybackslash}m{#1}} % centered fixed-width column
\newcolumntype{Y}{>{\centering\arraybackslash}X}        % centered X column

\newcommand{\paperhdr}[2]{%
  \makecell[c]{#1\\(\citeauthor{#2}\\\citeyear{#2})}%
}

\usepackage{hyperref}

\usepackage[preprint]{icml2026}

\usepackage{amsmath}
\usepackage{amssymb}
\usepackage{mathtools}
\usepackage{amsthm}

\usepackage[capitalize,noabbrev]{cleveref}

\graphicspath{{Figures/}}

\theoremstyle{plain}

\theoremstyle{definition}

\theoremstyle{remark}

\usepackage[textsize=tiny]{todonotes}

\icmltitlerunning{Position: 3D Gaussian Splatting Watermarking Should Be Scenario-Driven and Threat-Model Explicit}

\begin{document}

\twocolumn[
  \icmltitle{Position: 3D Gaussian Splatting Watermarking \\Should Be Scenario-Driven and Threat-Model Explicit
}



  \icmlsetsymbol{equal}{*}

  \begin{icmlauthorlist}
    \icmlauthor{Yangfan Deng}{Uni}
    \icmlauthor{Anirudh Nakra}{Uni}
    \icmlauthor{Min Wu}{Uni}
  \end{icmlauthorlist}

  \icmlaffiliation{Uni}{Department of Electrical and Computer Engineering, University of Maryland, College Park, MD, United States}

  \icmlcorrespondingauthor{Yangfan Deng}{yfandeng@umd.edu}
  \icmlcorrespondingauthor{Min Wu}{minwu@umd.edu}

  \icmlkeywords{Machine Learning, ICML}

  \vskip 0.3in
]



\printAffiliationsAndNotice{}  % no special notice (required even if empty)

% ===== Part I =====
\begin{abstract}

3D content acquisition and creation are expanding rapidly in the new era of machine learning and AI. 3D Gaussian Splatting (3DGS) has become a promising high-fidelity and real-time representation for 3D content. Similar to the initial wave of digital audio-visual content at the turn of the millennium, the demand for intellectual property protection is also increasing, since explicit and editable 3D parameterization makes unauthorized use and dissemination easier. In this position paper, we argue that effective progress in watermarking 3D assets requires articulated security objectives and realistic threat models, incorporating the lessons learned from digital audio-visual asset protection over the past decades. To address this gap in security specification and evaluation, we advocate a scenario-driven formulation, in which adversarial capabilities are formalized through a security model. Based on this formulation, we construct a reference framework that organizes existing methods and clarifies how specific design choices map to corresponding adversarial assumptions. Within this framework, we also examine a legacy spread-spectrum embedding scheme, characterizing its advantages and limitations and highlighting the important trade-offs it entails. Overall, this work aims to foster effective intellectual property protection for 3D assets.

\end{abstract}
\section{Introduction}

Recent advances in machine learning and AI are accelerating the full lifecycle of 3D content, from capture and reconstruction to creation and distribution.
Consumer-grade acquisition pipelines, neural rendering, and generative models are steadily reducing the cost of producing high-quality 3D content.
In parallel, real-time viewers and online marketplaces make it increasingly convenient to package, transfer, and repurpose 3D assets for downstream uses such as augmented and virtual reality (AR/VR)~\cite{dengel2022review}. Consequently, 3D assets are emerging as a central modality in the digital media economy, where they are routinely reproduced, edited, and redistributed at scale.

This trend has been observed repeatedly in digital media evolution over the past several decades. When high-quality digital audio, image, and video became easy to copy and distribute in the 1990s, intellectual property protection emerged as a practical necessity, and watermarking became one of the core technical tools being considered to support copyright management and tracing under redistribution~\cite{cox2008digital, stamm2013information}. Today, 3D assets face a similar trade-off. Modern 3D representations are editable, which improves usability but also makes misuse and unauthorized redistribution easier. Among all 3D representations, 3D Gaussian Splatting (3DGS) has quickly become a prominent representation for high-fidelity novel-view synthesis due to its explicit parameterization and real-time rendering quality~\cite{kerbl20233d}. Traditional 3D representations, such as mesh-based pipelines, often require heavy reconstruction and texturing. Point clouds may struggle to capture complex appearance details~\cite{wegen2024survey}. And neural radiance field (NeRF) methods may struggle to represent complex lighting~\cite{mildenhall2021nerf}. In comparison,  3DGS offers an attractive balance between visual quality, editability, and practical deployability. As a result, 3DGS models are increasingly packaged and shared as transferable assets in practical workflows~\cite{guo2025articulatedgs,peng2024rtg}. Throughout this paper, we use 3DGS as a concrete embodiment of this broader shift toward learnable and redistributable 3D media assets.

As 3D assets become widely shared and repurposed in real-world workflows fostered by advances in AI and machine learning, protecting them against unauthorized redistribution and misuse becomes increasingly important. Watermarking 3D assets is associated with security assumptions and attack surfaces that are often more intricate than those in traditional audio-visual media. In particular, protection spans two coupled domains, the model domain of the 3D representation and the rendering domain. The latter consists of 2D renditions from potentially many viewpoints, and an adversary may access either the 3D model or the rendered outputs. In addition, 3DGS models are frequently reparametrized and reordered during distribution and processing, which can undermine index-anchored protection mechanisms~\cite{bagdasarian20253dgs}. Geometry and appearance are also tightly coupled, and different parameters exhibit highly non-uniform perceptual sensitivity, which makes the choice of embedding domain central to imperceptibility and payload~\cite{xie2024mesongs}. Finally, real-time rendering interfaces enable querying rendered views, enabling adaptive attacks such as multi-view probing and optimization against the detector~\cite{cayre2005watermarking}. These properties call for explicit and rigorous security specifications, rather than evaluation regimes that implicitly assume static media and non-adaptive post-processing.

Despite the growing interest in 3DGS watermarking, a concerning pattern in the R\&D efforts of this area is the lack of systematic and articulated security objectives and realistic threat models. Many works leave adversarial access largely unspecified and unjustified. They also treat the setting as either model-level or rendering-output-level without defining a taxonomy for concrete deployment scenarios, and they omit key management. In addition, several systems adopt the assumption of making a watermark detector available, such as HiDDeN~\cite{zhu2018hidden}, overlooking lessons learned from the past literature and industry practices decades ago. These gaps make it difficult to compare methods under shared assumptions and leave security claims hard to reproduce, validate, or justify.

% In this position paper, we argue that effective progress in watermarking 3D assets requires scenario-grounded security objectives and threat models, incorporating lessons learned from digital audio-visual asset protection over the past decades~\cite{cox2002digital}. To address the gap in security specification and evaluation, we advocate a scenario-driven formulation in which adversarial capabilities are formalized through a security model. Based on this formulation, we construct a reference framework that organizes existing 3DGS watermarking methods and clarifies how design choices, such as keying, detector publicity, and embedding domains, map to corresponding adversarial assumptions. Within this framework, we also examine a legacy spread-spectrum embedding scheme, characterizing its advantages and limitations and highlighting the important trade-offs it entails. Overall, our goal is to foster qualified and reproducible intellectual property protection for 3D assets, while emphasizing that no single watermarking framework can be optimal across all deployment scenarios.

Noting these security weaknesses and limitations, we advocate a scenario-driven blueprint for 3DGS watermarking where the threat model should be defined by the deployment scenario rather than by the media type alone. First of all, the application and security objectives should be explicitly stated, since the success criteria, acceptable costs, and tradeoffs can vary significantly and are closely associated with the needs of practical applications. Second, the threat model should be made explicit within the chosen scenario. By using an access vector to formalize what the adversary can access and what capabilities it has, the threat levels can be described as subsets or clusters of access vectors to enable systematic evaluations and quantitative comparisons. Third, algorithm design should match the selected threat model, with particular attention to keying, detector availability, and the implied security boundary. Evaluation should follow a unified protocol that measures both effectiveness and risk. \textbf{We emphasize that no single watermarking framework can be perfect for all scenarios. Without scenario-grounded definitions, watermarking applications can easily collapse into engineering-oriented data hiding and fail to address the core security questions or achieve the intended protection objective.}

The remainder of this paper is organized as follows.
Section~2 provides preliminaries on 3DGS and the definitions for threat modeling. Section~3 analyzes important scenarios with the defined security system, with a focus on forensic watermarking. Section~4 builds a reference system and reviews representative methods with a structured discussion. Section~5 presents a spread-spectrum baseline allowing for explicit employment of a security key and experimental results to illustrate the important trade-offs and challenges. Section~6 presents the call to action, outlining critical open questions for the technical community to investigate. Section~7 concludes the paper.

\section{Preliminaries and Notation}
\label{sec:preliminaries-notation}
\begin{figure*}[t]
  \centering
  \includegraphics[width=\textwidth]{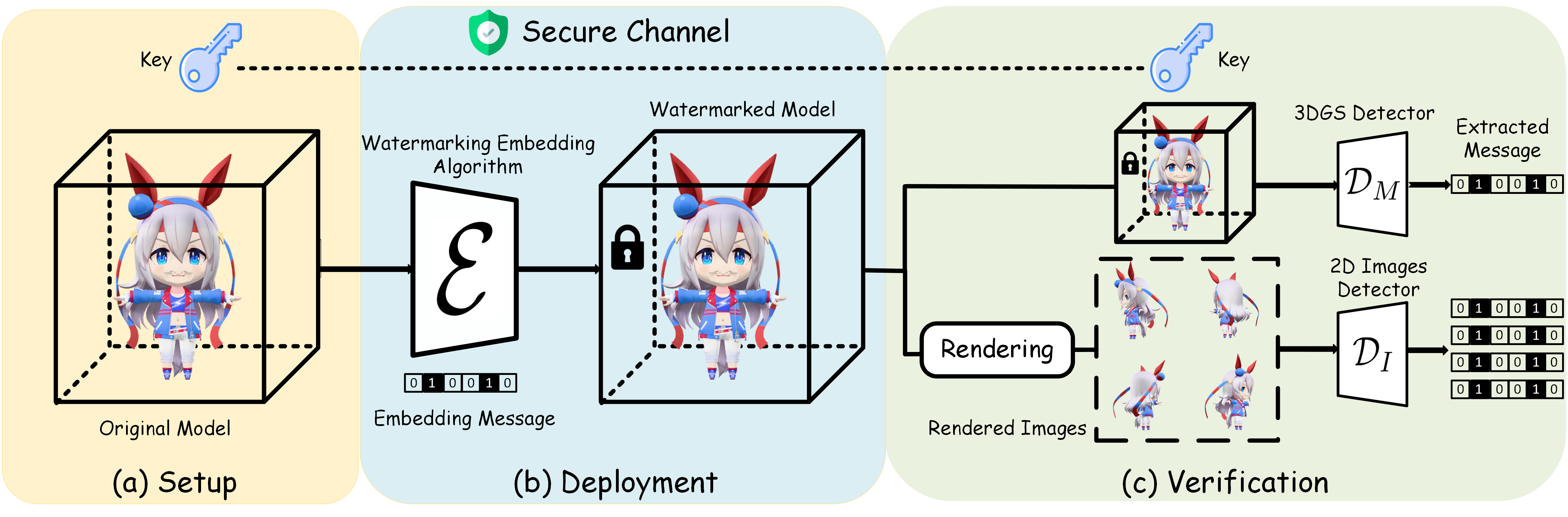}
  \caption{An authenticated watermarking system. It consists of three parts, \textbf{setup}, \textbf{embedding}, and \textbf{verification}. It is essential to ensure that the key is transmitted directly from the setup side to the verification portal via a secure channel.}
  \label{fig: The authenticated watermarking system}
\end{figure*}

\paragraph{3D Gaussian Splatting.} The primitive of 3DGS is referred to as a Gaussian, denoted as $g_i$~\cite{kerbl20233d}. Accordingly, a 3DGS scene $\mathcal{M}$ is represented as a set of anisotropic 3D Gaussians:
\begin{equation}
\label{eq:gs-model}
\mathcal{M}=\{g_i\}_{i=1}^{N},\qquad
g_i=\{\mu_i,\Sigma_i,d_i,f_i\},
\end{equation}
where $\mu_i\in\mathbb{R}^{3}$ specifies the coordinates of the Gaussian center, $\Sigma_i\in \mathrm{SPD}(3)$ is a positive definite covariance matrix encoding the anisotropic shape, $d_i\in[0,1]$ denotes the opacity, and $f_i\in\mathbb{R}^{d}$ parameterizes the spherical-harmonic coefficients for view-dependent color. For any 3D location $x\in\mathbb{R}^{3}$, the unnormalized density of the $i$-th Gaussian is
\begin{equation}
\label{eq:3d-gaussian}
\mathcal{G}_i(x)
=\exp\!\Bigl(-\tfrac{1}{2}(x-\mu_i)^\top \Sigma_i^{-1}(x-\mu_i)\Bigr).
\end{equation}

Given a viewpoint $v$, the renderer maps the 3DGS scene $\mathcal{M}$ to a rendered image $I_v$. For a pixel $p$ on the image plane, $\mathcal{N}(p)$ denotes the set of Gaussians whose splats overlap $p$. Each Gaussian $g_i$ contributes an RGB color $c_i(p)\in\mathbb{R}^3$ and an opacity weight $\alpha_i(p)\in[0,1]$. The rendered pixel color is obtained by standard front-to-back compositing:
\begin{equation}
\label{eq:alpha-compositing-simple}
C(p) \;=\; \sum_{i\in \mathcal{N}(p)} c_i(p)\,\alpha_i(p)
\prod_{\substack{j\in \mathcal{N}(p),\, j \prec i}}\bigl(1-\alpha_j(p)\bigr),
\end{equation}
where $j \prec i$ represents that splat $j$ is in front of splat $i$ along the viewing ray corresponding to pixel $p$.
Equivalently, defining the transmittance
\begin{equation}
\label{eq:transmittance-simple}
T_i(p) \;=\; \prod_{\substack{j\in \mathcal{N}(p),\, j \prec i}}\bigl(1-\alpha_j(p)\bigr),
\end{equation}
then we have $C(p)=\sum\limits_{i\in\mathcal{N}(p)} c_i(p)\,\alpha_i(p)\,T_i(p)$.
Finally, the overall rendering process is represented as
\begin{equation}
\label{eq:renderer-simple}
I_v \;=\; \mathcal{R}(\mathcal{M}, v;\theta_R),
\end{equation}
where $\mathcal{R}$ is the renderer,  $v$ represents the viewpoint, and $\theta_R$ denotes the parameters of the renderer.

\paragraph{The Authenticated Watermarking System.}
As shown in Figure~1, an authenticated watermarking system for 3DGS generally consists of three stages: \textbf{setup}, \textbf{embedding}, and \textbf{verification}~\cite{wu2003data1, wu2003multimedia}.

In the setup stage, the system specifiesa cryptographically secure key $K$ and its usage, such as how $K$ will be invoked during deployment. And a copy of $K$ is stored at the verification portal via a secure channel. Meanwhile, the system sets up the watermarking signal detectors and the rendering viewer at the verification portal. In particular, two types of detectors should be considered, a 3DGS model-level detector $\mathcal{D}_M$ for 3DGS models and an image/video-level detector $\mathcal{D}_I$ for rendered images and videos with $I_v$~\cite{zhao2024invisible, muller2025black}.

In the embedding stage, the watermark embedder $\mathcal{E}$ inserts a message $m\in\{0,1\}^L$ into an original 3DGS model $\mathcal{M}$ to produce a watermarked model $\mathcal{M}_w$:
\begin{equation}
\label{eq:embedder}
\begin{aligned}
\mathcal{E}(\mathcal{M}, m, K;\theta_E) = \mathcal{M}_w
:= \{g_i^{(w)}\}_{i=1}^{N_w}, \\
% g_i^{(w)} = \{\mu_i^{(w)},\Sigma_i^{(w)},d_i^{(w)},f_i^{(w)}\},
\end{aligned}
\end{equation}
where $\theta_E$ are the parameters of the embedding algorithm. Generally, $N_w \leq N$ since some watermarking algorithms may prune the set of Gaussians~\cite{jang20253d}.

In verification, an independent verification portal provides limited query access for watermark verification~\cite{cayre2005watermarking,quiring2018adversarial}. Given a suspect 3D model or rendered images, it can output a detection decision and the extracted watermarking message:
\begin{equation}
\label{eq:detectors}
\begin{aligned}
\mathcal{D}_M(\mathcal{M}_w, K;\theta_D)=(\hat y,\hat m,S),\\
\mathcal{D}_I(\{I_{v_t}\}_{t=1}^{T}, K;\theta_D)=(\hat y,\hat m,S),
\end{aligned}
\end{equation}
where $\theta_D$ denotes the parameters of the detector, $\hat y\in\{0,1\}$ indicates watermark presence, $\hat m$ is the extracted watermarking message, and $S$ denotes evaluation metrics, such as PSNR and SSIM. The portal is not public and only grants authenticated users a limited number of verification queries to prevent adaptive query attacks~\cite{cox1997secure2}.

\paragraph{Security Evaluation Model.}
Given the 3DGS representation and the authenticated watermarking system, we define the following access vector to characterize the privileges of an adversary:
$\mathcal{A}=\bigl[
\mathrm{Access}\text{-}\mathcal{M},\allowbreak\ 
\mathrm{Access}\text{-}\mathcal{M}_w,\allowbreak\ 
\mathrm{Access}\text{-}\mathcal{E},\allowbreak\ 
\mathrm{Access}\text{-}\mathcal{D},\allowbreak\ 
\mathrm{Oracle}\text{-}\mathcal{D},\allowbreak\ 
\mathrm{Oracle}\text{-}\mathcal{R},\allowbreak\ 
\mathrm{Key}\text{-}K
\bigr],$ where Access indicates full access to the corresponding artifact, including model files or implementation details, and Oracle indicates query access provided by the verification portal. For each entry, the value is set to $1$ if the adversary has the specified full or oracle access, and to $0$ otherwise. Unless otherwise specified, the secret key is unknown to the adversary and the original model is not directly accessible, which means that
$\mathrm{Key}\text{-}K=\mathrm{Access}\text{-}\mathcal{M}=0$.

Based on the formulation of access vector $\mathcal{A}$, we further categorize adversarial settings into three box regimes: black/grey/white box. It can reflect different risk levels and is instantiated by specific patterns of entries in $\mathcal{A}$. The access vectors corresponding to the three box regimes are denoted as $\mathcal{A}_{\mathrm{bb}}$, $\mathcal{A}_{\mathrm{gb}}$, and $\mathcal{A}_{\mathrm{wb}}$.
In the black-box regime, the adversary has no direct access to internal artifacts and can only interact with the system through oracle queries. Classical attacks include submitting inputs to the rendering oracle $\mathrm{Oracle}\text{-}\mathcal{R}$ to obtain rendered images or videos. These access vectors are typically in the form of: $\mathcal{A}_{\mathrm{bb}}=\bigl[
\mathrm{Access}\text{-}\mathcal{M}=0,\allowbreak\ 
\mathrm{Access}\text{-}\mathcal{M}_w=0,\allowbreak\ 
\mathrm{Access}\text{-}\mathcal{E}=0,\allowbreak\ 
\mathrm{Access}\text{-}\mathcal{D}=0,\allowbreak\ 
\mathrm{Oracle}\text{-}\mathcal{D},\allowbreak\ 
\mathrm{Oracle}\text{-}\mathcal{R},\allowbreak\ 
\mathrm{Key}\text{-}K=0
\bigr]$.

% \begin{equation}
% \label{eq:black-box}
% \begin{aligned}
% \mathcal{A}_{\mathrm{bb}}:\quad
% &\mathrm{Oracle}\text{-}\mathcal{R}=1\ \text{and/or}\ \mathrm{Oracle}\text{-}\mathcal{D}=1,\\
% &\mathrm{Access}\text{-}\mathcal{M}
% =\mathrm{Access}\text{-}\mathcal{M}_w
% =\mathrm{Access}\text{-}\mathcal{E}\\
% &=\mathrm{Access}\text{-}\mathcal{D}
% =\mathrm{Key}\text{-}K
% =0.
% \end{aligned}
% \end{equation}
In the white-box regime, the adversary can fully access the watermarked model and the details of the embedding and detection algorithms, while the original model and the key remain unavailable. Accordingly, we have:
% \begin{equation}
% \label{eq:access-vector}
% \begin{aligned}
% \mathcal{A}_{\mathrm{wb}}=\bigl(
% &\mathrm{Access}\text{-}\mathcal{M}=0,\ 
% \mathrm{Access}\text{-}\mathcal{M}_w=1,\ 
% \mathrm{Access}\text{-}\mathcal{E}=1,\\
% &\mathrm{Access}\text{-}\mathcal{D}=1,\ 
% \mathrm{Oracle}\text{-}\mathcal{D},\ 
% \mathrm{Oracle}\text{-}\mathcal{R},\\ 
% &\mathrm{Key}\text{-}K=0
% \bigr),
% \end{aligned}
% \end{equation}
$\mathcal{A}_{\mathrm{wb}}=\bigl[
\mathrm{Access}\text{-}\mathcal{M}=0,\allowbreak\ 
\mathrm{Access}\text{-}\mathcal{M}_w=1,\allowbreak\ 
\mathrm{Access}\text{-}\mathcal{E}=1,\allowbreak\ 
\mathrm{Access}\text{-}\mathcal{D}=1,\allowbreak\ 
\mathrm{Oracle}\text{-}\mathcal{D},\allowbreak\ 
\mathrm{Oracle}\text{-}\mathcal{R},\allowbreak\ 
\mathrm{Key}\text{-}K=0
\bigr]$,
% \begin{equation}
% \label{eq:white-box}
% \begin{aligned}
% \mathcal{A}_{\mathrm{wb}}:\quad
% &\mathrm{Access}\text{-}\mathcal{M}=0,\ 
% \mathrm{Access}\text{-}\mathcal{M}_w=1,\\
% &\mathrm{Access}\text{-}\mathcal{E}=1,\ 
% \mathrm{Access}\text{-}\mathcal{D}=1,\\
% &\mathrm{Key}\text{-}K=0.
% \end{aligned}
% \end{equation}
where the oracle entries $\mathrm{Oracle}\text{-}\mathcal{R}$ and $\mathrm{Oracle}\text{-}\mathcal{D}$ may additionally be $0$ or $1$ depending on the deployment.

The grey-box regime refers to any intermediate access pattern that is strictly stronger than pure oracle access but does not reach the full-access conditions of the white-box regime. Then, the grey-box settings satisfy
% \begin{equation}
% \label{eq:grey-box}
% \mathcal{A}_{\mathrm{gb}}:\quad
% \mathcal{A}_{\mathrm{bb}} \prec \mathcal{A}_{\mathrm{gb}} \prec \mathcal{A}_{\mathrm{wb}}.
% \end{equation}
$\mathcal{A}_{\mathrm{gb}}:\allowbreak\quad
\mathcal{A}_{\mathrm{bb}} \prec \mathcal{A}_{\mathrm{gb}} \prec \mathcal{A}_{\mathrm{wb}}$.
We group all the potential attacks for 3DGS into three levels according to where the adversary operates.\\
\noindent(i) \textbf{3DGS-level attacks}: the adversary directly manipulates the 3DGS representation, such as editing Gaussian parameters, pruning, resampling~\cite{chen2025guardsplat, jang20253d, huang2024gaussianmarker}.\\
\noindent(ii) \textbf{Image/video-level attacks}: the adversary only accesses rendered outputs and post-processes them, such as compression, cropping, resizing, frame dropping, re-encoding, and screen-recording~\cite{liang2025screenmark}.\\
\noindent(iii) \textbf{Neural-network-level attacks:} the adversary exploits a public or queryable detector and uses learning optimization to erase or spoof watermarking signals, such as gradient-based removal, surrogate modeling, and adversarial optimization~\cite{cheng2024badpart, liu2025stealthy}.

\section{Important Use-Case Scenario: Embedded Fingerprinting / Forensic Watermarking}
\label{subsec:scenario-b-fingerprinting-leakage-tracing-accountability}
To motivate scenario-driven modeling for 3DGS watermarking, we shall delineate deployment settings. While watermarks have been proposed to facilitate tampering detection and copyright management~\cite{hartung2002multimedia, wu2003multimedia}. Embedded fingerprinting, also known as Hollywood forensic watermarking, enables tracing individual copies of audio-visual assets and remains the primary successful real-world watermarking application with sustained, broad deployment~\cite{munoz2004suspected, movielabs_ecp_v1_4}. Thus, we use forensic watermarking as the reference setting for security and adversary analysis.

Another typical scenario, Copyright Ownership Verification, is discussed in \hyperref[Appendix_A]{Appendix~A}. Here, the watermark serves as a copyright indicator. For each scenario, we (i) define the overall access vector and the primary threat models, (ii) specify representative sub-scenarios with access vectors under black/grey/white-box regimes, respectively, and (iii) summarize keying mechanisms from traditional media that are effective under analogous assumptions.

% Fingerprinting for leakage tracing and accountability refers to the setting where a model is distributed to multiple recipients or collaborators, and each delivered copy is embedded with a distinct, identity-bound fingerprint~\cite{tardos2008optimal, trappe2003anti}.
% When unauthorized redistribution or piracy occurs, the fingerprint can be recovered from the intercepted 3DGS model or its rendered outputs.
% This recovery enables identification of the leakage source and supports attribution and accountability~\cite{chor1994tracing}.
% The main goal of this scenario is not to prove ownership, but to answer who leaked the content.

\subsection{Black-box Regime}
\label{subsubsec:per-buyer-fingerprinting-personalized-distribution}
The access vector for this regime can be summarized as
$\mathcal{A}=[0,0,0,0,\mathrm{Oracle}\text{-}\mathcal{D},\mathrm{Oracle}\text{-}\mathcal{R},0]$,
with $\mathrm{Oracle}\text{-}\mathcal{D},\mathrm{Oracle}\text{-}\mathcal{R}\in\{0,1\}$.
In this setting, attacks rely on observable outputs, namely the rendered images or videos, and interactions with online rendering or tracking interfaces.

The common sub-scenarios in this regime include:

\begin{enumerate}[label=(\roman*), leftmargin=20pt, itemsep=-3pt, topsep=-3pt]
\item Cloud restreaming: this refers to capturing rendered frames from a 3D asset or rebroadcasting screen recordings of the rendering, whereby attribution of the owner or copyright information is intentionally blurred. Its access vector is $[0,0,0,0,0,1,0]$.

\item Passive leakage of the rendered fixed-view images/videos: this means that only a fixed-trajectory video or a small set of screenshots is available. The corresponding access vector is $[0,0,0,0,0,0,0]$, representing the least powerful adversary setting for fingerprinting.

\item Using a tracing portal: this refers to a setting where processed excerpts, such as rendered visual segments, possibly after compression and/or cropping, are repeatedly submitted to an available detector. The returned portal feedback is then used in subsequent iterative probing until the forensic watermark may be attenuated so that its tracing outcome becomes unreliable. The corresponding access vector is $[0,0,0,0,1,0,0]$.
\end{enumerate}

In 3D assets, the verifier often observes only rendered outputs, either a multi-frame video along a camera trajectory or a set of views. Because watermark evidence in the rendering domain can be weak or view-dependent, keyed aggregation enables accumulating per-frame or segment statistics into a confidence-scored suspect set~\cite{cox1997secure2}. In addition, the key can be used in a challenge-response way to determine what viewpoints to render or what to select from a server-specified set of viewpoints. Therefore, stable attribution is required under multi-view and multi-segment challenges, which substantially increases the cost of evasion~\cite{cayre2005watermarking}.

\begin{table*}[t]
\centering
\caption{Summary of existing 3DGS watermarking methods: embedding design and experimental setup.}
\label{tab:3dgs_wm_summary_merged}
\setlength{\tabcolsep}{5pt}
\renewcommand{\arraystretch}{1.25}
\scriptsize

\begin{tabularx}{\textwidth}{|
C{2.55cm}|  % Papers (wider)
C{2.05cm}|  % Bit String
C{2.25cm}|  % Embedding Domain
C{2.45cm}|  % Embedding Stage (wider)
C{1.75cm}|  % Detector
Y|          % Datasets (flex + centered)
C{1.75cm}|} % GPU
\hline
Papers &
Bit String &
Embedding Domain &
Embedding Stage &
Detector &
Datasets &
GPU \\
\hline

GuardSplat~\cite{chen2025guardsplat} &
\makecell[c]{16 / 32 / 48 bits} &
\makecell[c]{SH Parameters} &
\makecell[c]{Post-hoc Optimization} &
\makecell[c]{Private} &
\makecell[c]{Blender\\LLFF} &
\makecell[c]{Single RTX 3090} \\
\hline

3D-GSW~\cite{jang20253d} &
\makecell[c]{32 / 48 / 64 bits} &
\makecell[c]{Global} &
\makecell[c]{Post-hoc Optimization} &
\makecell[c]{Public} &
\makecell[c]{Blender\\LLFF\\Mip-NeRF~360} &
\makecell[c]{Single A100} \\
\hline

GS-Marker~\cite{li2025gs} &
\makecell[c]{32 bits} &
\makecell[c]{Global} &
\makecell[c]{Single Forward\\Process} &
\makecell[c]{Public} &
\makecell[c]{Objaverse\\Blender\\OmniObject\allowbreak 3D} &
\makecell[c]{8 V100} \\
\hline

MarkSplatter~\cite{huang2025marksplatter} &
\makecell[c]{32 / 48 bits} &
\makecell[c]{Splatter Images} &
\makecell[c]{Single Forward\\Process} &
\makecell[c]{Private} &
\makecell[c]{Objaverse\\Google Scanned\\Objects} &
\makecell[c]{8 V100} \\
\hline

Water-GS~\cite{tan2024water} &
\makecell[c]{48 bits} &
\makecell[c]{Global} &
\makecell[c]{Post-hoc Optimization} &
\makecell[c]{Public} &
\makecell[c]{Blender\\LLFF\\Tanks\&Temples} &
\makecell[c]{Single RTX 3090} \\
\hline

Gaussian\allowbreak Marker~\cite{huang2024gaussianmarker} &
\makecell[c]{48 bits} &
\makecell[c]{Global} &
\makecell[c]{Post-hoc Optimization} &
\makecell[c]{Public} &
\makecell[c]{Blender\\LLFF\\Mip-NeRF~360} &
\makecell[c]{Single V100} \\
\hline

\end{tabularx}
\end{table*}

% =========================
% The table
% =========================
\begin{table*}[t]
\centering
\caption{Threat-model coverage and evaluation metrics of recent 3DGS watermarking methods.}
\label{tab:threat_model_matrix_3dgs}
\setlength{\tabcolsep}{5pt}
\renewcommand{\arraystretch}{1.25}
\scriptsize

\begin{tabularx}{\textwidth}{|
>{\centering\arraybackslash}m{1.9cm}|
>{\centering\arraybackslash}m{3.4cm}|
*{6}{>{\centering\arraybackslash}X|}}
\hline

\multicolumn{2}{|c|}{Papers} &
\paperhdr{GuardSplat}{chen2025guardsplat} &
\paperhdr{3D-GSW}{jang20253d} &
\paperhdr{GS-Marker}{li2025gs} &
\paperhdr{MarkSplatter}{huang2025marksplatter} &
\paperhdr{Water-GS}{tan2024water} &
\paperhdr{GaussianMarker}{huang2024gaussianmarker} \\

\hline

\multicolumn{2}{|c|}{Attacking Domains} &
Rendered Images &
Rendered Images + 3DGS model &
Rendered Images + 3DGS model &
Rendered Images + 3DGS model &
3DGS model &
Rendered Images + 3DGS model \\
\hline

% --- THREAT MODELS block (multirow spans BOTH 2D+3D parts) ---
\multirow{14}{*}{\makecell[c]{Threat\\Models}}
& Gaussian Noise ($\sigma=0.1$)
    & \cmark & \cmark & \cmark & \cmark &        & \cmark \\
\cline{2-8}
& Rotation ($\pm\pi/6$)
    & \cmark & \cmark & \cmark & \cmark &        &        \\
\cline{2-8}
& Scaling ($75\%$)
    & \cmark & \cmark &        & \cmark &        & \cmark \\
\cline{2-8}
& Gaussian blur ($\sigma=0.1$)
    & \cmark & \cmark &        & \cmark &        & \cmark \\
\cline{2-8}
& 2D Cropping (40\%)
    & \cmark & \cmark & \cmark & \cmark &        &        \\
\cline{2-8}
& Brightness ($0.5 \sim 1.5$)
    & \cmark &        &        &        &        &        \\
\cline{2-8}
& JPEG Compression ($Q=50\%$)
    & \cmark & \cmark & \cmark & \cmark &        & \cmark \\
\cline{2-8}
& Translation ($20\%$)
    &        &        &        & \cmark &        &        \\
\cline{2-8}
& VAE attack
    & \cmark &        &        &        &        &        \\
\cline{2-8}
& \textit{Gaussian noise ($\sigma=0.1$)}
    &        &        &        & \cmark & \cmark & \cmark \\
% ---- IMPORTANT: separator between Noise(3D) and Dropout(3D)
%      DO NOT draw across col 1; hence \cline{2-8} (no \hline here)
\cline{2-8}
& \textit{Dropout ($20\%$)}
    &        & \cmark & \cmark & \cmark & \cmark &        \\
\cline{2-8}
& \textit{3D Cropping ($0.5$)}
    &   & \cmark & \cmark & \cmark & \cmark & \cmark \\
\cline{2-8}
& \textit{Cloning ($20\%$)}
    &        & \cmark &        &        &        &        \\
\cline{2-8}
& \textit{Translation ($20\%$)}
    &        &        &        & \cmark &        & \cmark \\
\hline

\multicolumn{2}{|c|}{Evaluation Metrics} &
\makecell{Bit Accuracy\\Fidelity\\Efficiency} &
\makecell{Bit Accuracy\\Fidelity\\Efficiency} &
\makecell{Bit Accuracy\\Fidelity\\Efficiency} &
\makecell{Bit Accuracy\\Fidelity\\Efficiency} &
\makecell{Bit Accuracy\\Fidelity\\Efficiency} &
\makecell{Bit Accuracy\\Fidelity\\Model Distortion} \\
\hline

\end{tabularx}
\end{table*}

\subsection{White-box Regime}
Under the white-box regime for forensic watermarking, the access vector is $\mathcal{A}=[0,1,1,1,1,1,0]$.
In this setting, the adversary has the internal implementations and the weights of the detector.

Once the detection mechanism and configuration are known, evasion can be performed by directly optimizing against the tracing objective with minimal distortion.
Samples can even be constructed to falsely implicate an innocent party~\cite{memon2001buyer}.
Compared with ownership verification, fingerprinting is more sensitive to the legal and security consequences of false accusation and framing.
Therefore, the use of a key must remain the root of trust.
Keyed subset selection, projection, permutation, and thresholds are used so that access to the weights does not translate into reliable evasion~\cite{cox1997secure2}. In addition, signatures or message authentication codes can be used to bind claims and reduce the risk of framing~\cite{rivest1978method}. In practice, the objective is set to emphasize a strong evidentiary chain, a high cost of framing, and auditable re-verification~\cite{memon2001buyer}.

\subsection{Grey-box Regime}
\label{subsubsec:collusion-attacks-collusion-resilient-requirements}
In this grey-box regime, the access vector is expressed as
$\mathcal{A}=[0,1,0,0,\mathrm{Oracle}\text{-}\mathcal{D},\mathrm{Oracle}\text{-}\mathcal{R},0]$
with $\mathrm{Oracle}\text{-}\mathcal{D},\mathrm{Oracle}\text{-}\mathcal{R}\in\{0,1\}$.
One or more multiple personalized copies may be obtained by the adversary, enabling model-level editing, re-optimization, and splicing or mixing across copies.
As a result, most threat models focus on the 3DGS model level.

The offline forensic tracing and oracle-guided model evasion scenarios are typical cases:

\begin{enumerate}[label=(\roman*), leftmargin=20pt, itemsep=-3pt, topsep=-3pt, parsep=5pt, partopsep=0pt]
\item Offline forensic tracing includes the ability to track legitimate recipients, supply-chain or collaborator leaks, and collusive reconstruction. The access vector is $[0,1,1,0,0,1,0]$.

\item Oracle-guided model evasion means that model manipulation is combined with repeated queries to a tracing interface, and the feedback from the watermark detector in the oracle is treated as a cost criterion to support iterative fingerprint removal. The corresponding access vector is $[0,1,1,0,1,0,0]$.
\end{enumerate}

For this regime, the key is commonly used for individualized codeword generation, randomized embedding, or tracing rules~\cite{tardos2008optimal}.
This design enables the intended recipient as the source of the unintended distribution to be decoded from the leaked copy of the content~\cite{chor1994tracing}.
It also supports collusion resilience by returning at least one traitor or a small suspect set~\cite{trappe2003anti}.
Adversaries’ effectiveness through iterative updates to estimate and remove the watermark via repeated queries of the watermark detector can be further reduced by introducing keyed challenges, truncating portal outputs, and auditing and limiting the number of queries.

% ===== Part II =====
\section{Alternative Views}
\label{sec:overview-design-principles-reference-system}
\subsection{View 1: Engineering-Oriented Robustness Benchmarks Are Sufficient}

This view argues that progress in 3DGS watermarking is driven by standardized benchmarks on robustness, rather than by developing security models and articulating deployment-specific assumptions. It advocates that the community should prioritize a shared suite of attacks and metrics across perturbations on both the 3DGS model and visual data. As shown in Table~1 and Table~2, such a suite allows methods to be compared directly and iterated quickly. Recent papers on 3DGS watermarking usually report the bit-level accuracy under common distortions, together with fidelity metrics such as PSNR, SSIM, and LPIPS~\cite{chen2025guardsplat, jang20253d, li2025gs, huang2025marksplatter}. These works imply that adopting consistent payload reporting, such as $32/48/64$ bits, and expanding benchmark coverage are effective ways for reproducibility~\cite{jang20253d}.

We would like to note that a limitation of this view is that the breadth of robustness benchmark does not necessarily capture security-relevant capabilities, especially when the adversary is able to access the detector and perform optimization to circumvent the watermarking. That said, robustness and security are not mutually exclusive, as expanding the set of robustness metrics considered is beneficial to serve as a prerequisite for security and to achieve clarity and reproducibility. Robustness alone is not sufficient for security. Simply enumerating more threat models does not by itself make a watermarking framework secure~\cite{craver2002resolving, cox1997secure2}. The implication is that a benchmark-first agenda may yield strong and comparable numbers, while protocol choices such as keying, detector availability, and query controls remain underspecified and vary across deployments, making the system vulnerable to attacks~\cite{quiring2018forgotten}.

\subsection{View 2: Public Detectors Enable Open Evaluation and Faster Progress}

This view argues that public detectors should be treated as a design choice, as they increase transparency and reproducibility. The proponents of this view advocate that benchmarking should be done without dependence on proprietary verification portals. In Table 1, several representative 3DGS watermarking methods rely on image-level detectors, such as HiDDeN, for watermark extraction from rendered outputs~\cite{jang20253d, li2025gs, tan2024water, huang2024gaussianmarker}. Public detectors also support stress testing. They enable different attacks to be implemented in a consistent and specified manner across papers, rather than being approximated through limited oracle access.

However, detector availability also reshapes the threat landscape as adversaries can exploit the detector to adaptively remove or insert watermarks. A real-world example is the Secure Digital Music Initiative (SDMI)~\cite{sdmi}, which exposed a public detection oracle to mimic a blind detector inside a digital music player. Despite the lack of unwatermarked references and the absence of disclosed procedures, the community repeatedly observed that public detector access allows attackers to collect input--output pairs and mount oracle attacks, often sufficing to evade detection even when the detector is complex~\cite{craver2001reading,wu2001analysis}. When the detector is public, attackers can replicate the pipeline and optimize against it through gradient-based attacks, enabling selective erasure and targeted counterfeiting~\cite{quiring2018forgotten, tondi2016smart}. Therefore, security-oriented deployment scenarios should carefully avoid reliance on a public detector. If a public detector is adopted for open evaluation, adaptive attacks should be explicitly included in the threat model, and use of the detector should be circumvented.

\begin{figure*}[t]
  \centering
  \includegraphics[width=\textwidth]{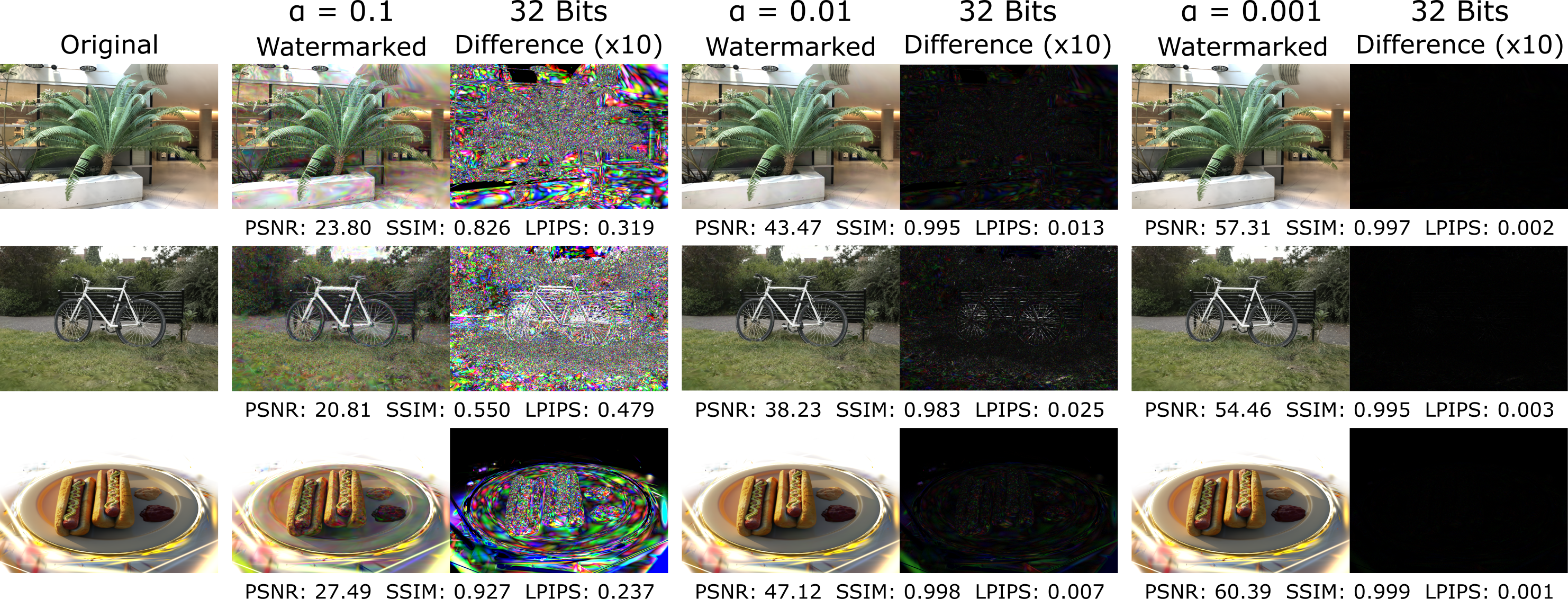}
  \caption{Qualitative fidelity under different spread-spectrum embedding amplitudes. We visualize rendered images of the original and watermarked 3DGS scenes, together with the corresponding difference images ($\times 10$), under payload $B=32$ and three embedding strengths $\alpha \in \{0.1,\,0.01,\,0.001\}$. Results are shown for three scenes, LLFF \emph{fern}, Mip-NeRF~360 \emph{bicycle}, and Blender \emph{hotdog}.}
  \label{fig:tradeoffA}
\end{figure*}

\subsection{View 3: Keying Is Optional in Learning-Based Watermarking}

This view argues that keying is not essential in learning-based watermarking when the detector is deployed as a private component~\cite{boenisch2021systematic}. Third parties cannot extract the watermark because the decoding function and its parameters are not publicly executable~\cite{cox1997secure2}. Operational overhead is reduced, as key generation, secure distribution, storage, and rotation are no longer required.

Without an explicit key, the watermarking scheme reduces to ``security by obscurity'', with the strength of protection hinging on keeping the detector confidential. This is contrary to a well-known security principle and makes both the threat model and the evaluation setup harder to define and reproduce~\cite{craver2002resolving}. If the detector weights are leaked or reverse engineered, the attacker can directly optimize against the detection objective~\cite{carlini2017towards}. Even without direct leakage, repeated interactions may allow adversaries to approximate the detector, enabling adaptive removal and forgery attacks. Consequently, if stronger security objectives are targeted, explicit keys and keyed mechanisms should be treated as necessary protocol components rather than optional implementation details.

% ===== Part III =====
% \begin{figure*}[t]
%   \centering
%   \includegraphics[width=\textwidth]{Trade-offA_2.png}
%   \caption{Qualitative fidelity with different payload lengths. Rendered original, watermarked, and difference ($\times 10$) images are represented with a fixed embedding strength $\alpha=0.01$ and payload lengths $B \in \{32,\,48,\,64\}$. The three rows correspond to LLFF \emph{fern}, Mip-NeRF~360 \emph{bicycle}, and Blender \emph{hotdog}, respectively.
% }
%   \label{fig:tradeoffA2}
% \end{figure*}

% ========= table (dataset column removed) =========
\begin{table*}[t]
\centering
\caption{Quantitative comparisons under different payload sizes and embedding amplitudes.}
\label{tab:amp_payload_fidelity}
\setlength{\tabcolsep}{6pt}
\renewcommand{\arraystretch}{1.15}
\begin{tabular}{l ccc ccc ccc}
\toprule
\multirow{2}{*}{Embedding Amplitude} &
\multicolumn{3}{c}{32 bits} &
\multicolumn{3}{c}{48 bits} &
\multicolumn{3}{c}{64 bits} \\
\cmidrule(lr){2-4}\cmidrule(lr){5-7}\cmidrule(lr){8-10}
& PSNR$\uparrow$ & SSIM$\uparrow$ & LPIPS$\downarrow$
& PSNR$\uparrow$ & SSIM$\uparrow$ & LPIPS$\downarrow$
& PSNR$\uparrow$ & SSIM$\uparrow$ & LPIPS$\downarrow$ \\
\midrule

$\alpha=0.1$   & 25.43 & 0.825 & 0.255 & 23.74 & 0.784 & 0.296 & 22.56 & 0.752 & 0.325 \\
$\alpha=0.01$  & 44.96 & 0.995 & 0.009 & 43.28 & 0.993 & 0.014 & 42.09 & 0.991 & 0.019 \\
$\alpha=0.001$ & 58.85 & 0.998 & 0.001 & 57.96 & 0.997 & 0.002 & 57.32 & 0.995 & 0.004 \\

\bottomrule
\end{tabular}
\end{table*}

\section{Classical Spread-Spectrum Embedding}
\label{sec:toy-embedding-algorithm}
Existing 3DGS watermarking works also lack a unified baseline that makes keying choices explicit and experiments reproducible. To address this gap, we apply the classical spread-spectrum watermarking baseline to the native 3DGS parameter space~\cite{cox1997secure2}. This formulation provides three concrete benefits:
(i) a transparent and fully reproducible embedding/detection pipeline;
(ii) a clearly defined key mechanism, including carrier selection, claim-to-message binding, and pseudo-random spreading code generation; and
(iii) two controllable trade-off protocols that enable subsequent work to be compared and iterated within a consistent coordinate system.

In particular, we represent the embedded message as a $B$-bit $\{\pm1\}$ payload vector and embed it via code-division multiplexing (CDM)~\cite{wu2003data2} over a reproducible transform-domain carrier pool. We instantiate this carrier pool directly in native 3DGS parameters and make its organization explicitly key-dependent. Therefore, the embedding layout is both unpredictable to an attacker and exactly reproducible for a verifier. A second key binds the claim to the payload, and a third key generates the pseudo-random spreading codes used for multi-bit superposition.

On the detection side, we follow a classical non-blind protocol. Given the original model and a suspect model, the detector forms a transform-domain residual over the same carrier pool and recovers each bit by correlation decoding followed by a sign test, reporting bit accuracy as the primary outcome. Moreover, we adopt two power-normalization strategies to expose the classical ``capacity--robustness--fidelity'' trade-offs on standard 3DGS benchmarks. Full implementation details are provided in \hyperref[Appendix_B]{Appendix~B}.

In the experiment, we evaluate the proposed spread-spectrum embedding and the non-blind detection baseline on three standard benchmarks, Blender~\cite{mildenhall2021nerf}, LLFF~\cite{mildenhall2019local}, and Mip-NeRF 360~\cite{barron2022mip}. Overviews of other commonly used datasets in this area, as well as detailed analyses of representative algorithms, are available in \hyperref[Appendix_C]{Appendix~C}. These datasets are widely adopted in the NeRF and 3DGS literature and cover both synthetic and real-world scene distributions. In addition, the common payload lengths $B\in\{32,48,64\}$ are considered. For the transform-domain processing, we set the mid-band interval $[0.10,0.18]$ as the candidate carrier range for the 1D DCT. The embedding strength parameter $\alpha$ is set to $\{0.1, 0.01, 0.001\}$, forming controlled groups with different embedding intensities.

For fidelity evaluation, we measure image quality on rendered images and utilize PSNR, SSIM, and LPIPS to evaluate the impact of watermark embedding on rendering fidelity. For robustness evaluation, detection experiments are conducted under a unified Gaussian-noise attack setting and use bit accuracy as the detection metric.

Figure~2 compares the original and watermarked rendered images together with the $\times 10$ difference images on three scenes with a fixed payload $B=32$ under different embedding amplitudes $\alpha$. As $\alpha$ decreases from $0.1$ to $0.01$ and $0.001$, visible artifacts and rendering residuals shrink, and the amplified differences transition from strong speckle patterns to localized residuals and then to almost black images. We also compare the fidelity with different payload lengths, which is displayed in Figure~4 in \hyperref[Appendix_D]{Appendix~D}.

Figure~3 examines the visual impact of different payload lengths while fixing the embedding strength at $\alpha=0.01$ under the fixed per-bit energy setting. While the renderings remain visually similar as the payload increases, the fidelity metrics degrade monotonically, with PSNR and SSIM decreasing and LPIPS increasing.

Table~3 reports fidelity metrics under different payload sizes and embedding amplitudes.  
The consistent monotonic trends indicate that this spread-spectrum scheme can still cause measurable fidelity degradation even when the rendering differences are not visually apparent.

\section{Call to Action}

We recommend adopting our scenario-driven blueprint as a \emph{minimum disclosure standard} for R\&D on 3DGS watermarking. First, each work should state the deployment scenario and security objective, together with a concrete success criterion and acceptable costs, since different objectives are associated with different trade-offs. Second, the threat model should be articulated within the chosen scenario, preferably including an access vector that formalizes what the adversary can access. Under this formulation, conventional regimes of white/black/gray boxes can be viewed as subsets of access patterns, enabling more comparable security statements. Third, the algorithm design should be aligned with the stated threat model and specify whether the scheme is keyed. Fourth, evaluation should follow a protocol that reports both effectiveness and exposure, rather than presenting engineering-oriented data hiding results under implicit default assumptions.

We recognize the significant challenge associated with cross-domain watermark detection, namely, to detect a watermark embedded at the 3DGS model level from the rendered images/videos. 3DGS watermarking spans two coupled domains, the model domain and the rendering domain. An ideal framework should account for both model-level and image/video-level detection. However, existing 3DGS watermarking pipelines rarely design detectors specifically for 2D rendered outputs. Instead, many works directly reuse public detectors from image watermarking, such as HiDDeN~\cite{zhu2018hidden}. While this reuse reduces engineering effort, as discussed in this paper, prior literature and past practices have shown that a public detector can become an entry point to be exploited by adversaries~\cite{cayre2005watermarking}. Therefore, we call on the community to embrace this research challenge, and design and clearly articulate framework-specific detectors that are tailored to the intended deployment objective and threat model, rather than defaulting to generic public detectors.

\begin{figure}[t]
  \centering
  \vspace{-0.4em}
  \includegraphics[width=\columnwidth]{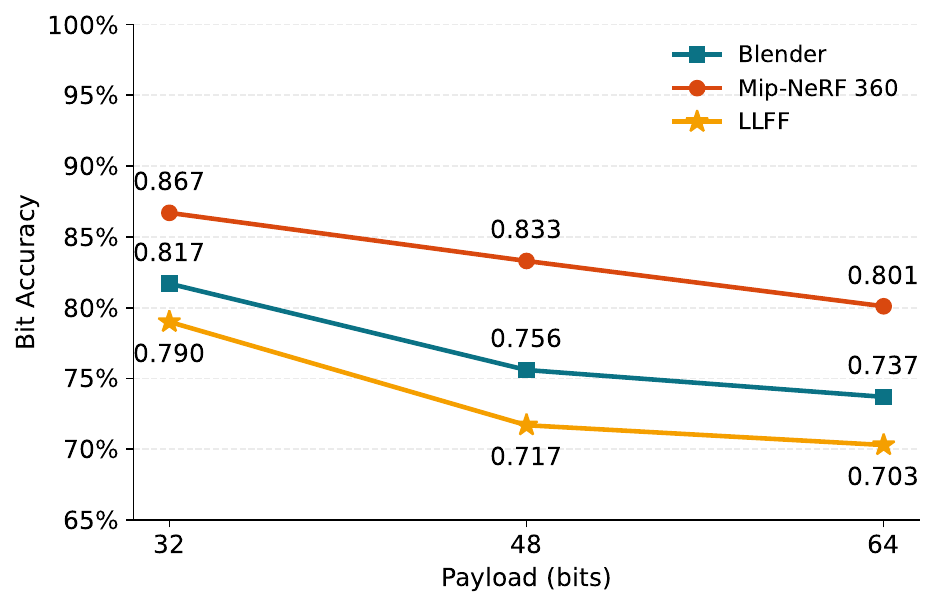}
  \vspace{-0.6em}
  \caption{The trade-off between bit accuracy and payload length. Bit accuracy is represented as a function of payload length $B \in \{32,\,48,\,64\}$ under a fixed-total embedding energy constraint.}
  \vspace{-0.8em}
  \label{fig:tradeoffB}
\end{figure}

\section{Conclusion}
\label{sec:conclusion}
In this position paper, we argue that the R\&D of 3DGS watermarking is more limited by unclear or questionable security definitions than by embedding techniques. We advocate a scenario-driven approach and utilize forensic watermarking scenarios as examples to demonstrate how to articulate the threat models and analyze the solutions. We incorporate the lessons learned from past literature and industry practices of audio-visual watermarking and construct a reference system to show how existing design choices implicitly assume choices of threat and adversary models. In addition, we provide a reproducible spread-spectrum baseline, in which keying mechanisms are made explicit and trade-off protocols are standardized. We hope these discussions help move 3DGS watermarking from ad-hoc data hiding toward scenario-grounded, comparable, and verifiable security.

\newpage


\bibliography{main}
\bibliographystyle{icml2026}

%%%%%%%%%%%%%%%%%%%%%%%%%%%%%%%%%%%%%%%%%%%%%%%%%%%%%%%%%%%%%%%%%%%%%%%%%%%%%%%
%%%%%%%%%%%%%%%%%%%%%%%%%%%%%%%%%%%%%%%%%%%%%%%%%%%%%%%%%%%%%%%%%%%%%%%%%%%%%%%
% APPENDIX
%%%%%%%%%%%%%%%%%%%%%%%%%%%%%%%%%%%%%%%%%%%%%%%%%%%%%%%%%%%%%%%%%%%%%%%%%%%%%%%
%%%%%%%%%%%%%%%%%%%%%%%%%%%%%%%%%%%%%%%%%%%%%%%%%%%%%%%%%%%%%%%%%%%%%%%%%%%%%%%
\newpage
\appendix
\onecolumn
\section{Additional Use-Case Scenario: Copyright Ownership Verification}
\label{Appendix_A}

Copyright ownership verification refers to the setting where an authorship or ownership dispute arises.
In this setting, the rights holder must demonstrate that a given 3DGS model or its rendered outputs indeed originate from the legitimate owner, thereby supporting ownership claims, enforcement, and licensing management.
The central objective of this scenario is a verifiable ownership statement.

\subsection{Black-box Regime}
\label{subsubsec:model-level-distribution-unauthorized-resale}
Under the black-box regime, the access vector can be summarized as $\mathcal{A}=[0,0,0,0,\mathrm{Oracle}\text{-}\mathcal{D},\mathrm{Oracle}\text{-}\mathcal{R},0]$,
with
$\mathrm{Oracle}\text{-}\mathcal{D},\mathrm{Oracle}\text{-}\mathcal{R}\in\{0,1\}$. In this section, the entry ordering of all access vectors follows that of Eq.~(8).
This setting indicates that the adversary has no access to $\mathcal{M}_w$ and the implementations or weights of the embedder and detector. Attacks are restricted to interactions with available oracle interfaces, which are mainly concentrated on the image/video-level~\cite{cox1997secure2}.

Two classical scenarios are considered, interactive viewers and broadcast media:

\begin{enumerate}[label=(\roman*), leftmargin=20pt, itemsep=-3pt, topsep=-3pt]
\item Interactive viewers means that the adversary can only capture screenshots or screen recordings from an interactive UI, while the 3DGS model file remains inaccessible. The corresponding access vector is $[0,0,0,0,0,1,0]$. The attack surface is primarily on the output side, such as compression artifacts and recording noise, in order to degrade watermark detectability.

\item Broadcast case is that only a fixed-trajectory video or a small set of images is available, without viewpoint control or repeated interaction. The corresponding access vector is $[0,0,0,0,0,0,0]$, which represents the weakest attack setting among the above.
\end{enumerate}

In these two scenarios, the key is commonly used for correlation-accumulation detection across multiple views or frames, where coherent aggregation of weak signals is enabled only with the correct key~\cite{hartung2002multimedia, cox1997secure2}.

\subsection{White-box Regime}
\label{subsubsec:portal-based-verification-non-selectable-viewpoints}
For the white-box regime, the access vector can be concluded as
$\mathcal{A}=[0,1,1,1,\mathrm{Oracle}\text{-}\mathcal{D},1,0]$,
with $\mathrm{Oracle}\text{-}\mathcal{D}\in\{0,1\}$.
This regime assumes that the adversary has access to the implementations and weights of both embedder and detector, enabling gradient-based attacks~\cite{carlini2017towards}.
Only $\mathrm{Key}\text{-}k$ and $\mathrm{Access}\text{-}\mathcal{M}$ remain unavailable.

Wo discuss two typical scenarios here, local white-box optimization and online white-box optimization:

\begin{enumerate}[label=(\roman*), leftmargin=20pt, itemsep=-3pt, topsep=-3pt]
\item Local white-box optimization corresponds to the setting where the adversary, after obtaining the weights of embedder and detector, directly performs gradient-based white-box optimization to modify $\mathcal{M}_w$, or its rendered images or videos, in order to erase the watermark with minimal distortion. The corresponding access vector is $[0,1,1,1,0,1,0]$.

\item Online white-box optimization corresponds to the setting where white-box backpropagation is available and a portal can also be queried to validate or align intermediate results, which improves attack efficiency. This access vector is $[0,1,1,1,1,1,0]$.
\end{enumerate}

White-box leakage is widely regarded as the most challenging setting in practice~\cite{quiring2018forgotten}. In this regime, the key is commonly treated as a means to raise the adversary's acquisition cost~\cite{cox1997secure2}. It serves as a runtime secret that controls where to look, how to read, and whether authentication is possible, through keyed selection, projection, and authentication~\cite{cox1997secure2}. Hence, leaking weights does not directly translate into precise erasure or reliable forgery.

\subsection{Grey-box Regime}
\label{subsubsec:render-only-release-passive-observation}
In the grey-box regime, the access vector can be summarized as
$\mathcal{A}=[0,1,0,0,\mathrm{Oracle}\text{-}\mathcal{D},\mathrm{Oracle}\text{-}\mathcal{R},0]$,
with $\mathrm{Oracle}\text{-}\mathcal{D},\mathrm{Oracle}\text{-}\mathcal{R}\in\{0,1\}$.
In this regime, the adversary can access $\mathcal{M}_w$ and therefore directly manipulate 3DGS parameters.
Meanwhile, $\mathrm{Access}\text{-}\mathcal{D}=0$ holds, and the detector weights remain inaccessible, which prevents white-box backpropagation through the detector. Accordingly, the main threat models concentrate at the 3DGS model level, including parameter perturbation, Gaussian pruning, and retraining or re-optimization.

Two canonical scenarios are considered in this regime, offline piracy with model resale and online piracy with model resale:

\begin{enumerate}[label=(\roman*), leftmargin=20pt, itemsep=-3pt, topsep=-3pt]
\item Offline piracy with model resale refers to the case where a pirate acquires $\mathcal{M}_w$ and performs offline modifications, such as pruning, re-optimization, or perturbation, aiming to suppress the watermark before resale. The corresponding access vector is $[0,1,1,0,0,1,0]$. Since $\mathrm{Oracle}\text{-}\mathcal{D}=0$ during the attack process, the attack is inherently blind.

\item Online piracy with model resale corresponds to the case where a pirate acquires $\mathcal{M}_w$ and can repeatedly query a verification portal, using the returned feedback to iteratively update the modified model until detection fails. The corresponding access vector is $[0,1,1,0,1,1,0]$.
\end{enumerate}

In both scenarios, the key is commonly used to determine the carrier subset and the bit mapping, through keyed carrier selection and keyed mapping~\cite{cox1997secure2}. This design makes precise erasure more difficult and forces broader model degradation in order to reliably remove the watermark~\cite{craver2002resolving}.

\section{The Details of Spread Spectrum Algorithm}
\label{Appendix_B}
\subsection{Keying Mechanism}
\label{subsec:carrier-selection-perturbation-injection}
Based on the Kerckhoffs assumption, the embedding algorithm should be treated as public, while security and reproducibility are governed by secret keys. Let the claim be a string $\mathcal{C}$ and the payload length be $B$. Three keys are introduced, $K_{\mathrm{sel}}$, $K_{\mathrm{code}}$, and $K_{\mathrm{seq}}$. The required randomness is derived from an HMAC-based random function $\mathrm{PRF}(K,\tau)\in\{0,1\}^{\ell}$. Using $K_{\mathrm{code}}$, the claim is mapped into a bipolar payload
\[
\mathbf{b}(\mathcal{C})=2\cdot \mathrm{PRF}(K_{\mathrm{code}}, \mathcal{C})_{1:B}-\mathbf{1}_\textit{B}\in\{-1,+1\}^{B},
\]
where $\mathrm{PRF}(\cdot)_{1:B}$ denotes taking the first $B$ bits. This construction makes it difficult to produce the correct payload for the same claim without $K_{\mathrm{code}}$, while ensuring exact reproducibility under the same key.

The carrier-organization key $K_{\mathrm{sel}}$ makes the embedding layout key-dependent and reproducible. Let the transform-domain candidate carrier pool be $\mathcal{P} = \{(p, k)\}$ with size $T$. The $K_{\mathrm{sel}}$ can generate a random permutation over carrier indices
\[
\pi \leftarrow \mathrm{Permute}\,\!\big(\mathrm{PRF}(K_{\mathrm{sel}},\,\mathcal{C}\,\Vert\,\texttt{perm}),\,T\big),
\]
where $\Vert$ denotes concatenation and \texttt{perm} is a fixed context tag used only to separate purposes. This step preserves the carrier set $\mathcal{P}$ and globally mixes carriers across fields and frequency indices, removing predictable structural ordering.

The spreading key generates a $T$-length random bipolar template for each bit position, used for both embedding superposition and correlation decoding. For each $j\in\{1,\ldots,B\}$, we define
\[
\mathbf{s}_j(\mathcal{C})=2\cdot \mathrm{PRF}(K_{\mathrm{seq}},\,\mathcal{C}\,\Vert\, j)_{1:T}-\mathbf{1_\textit{T}}\in\{-1,+1\}^{T}.
\]
The detector reproduces $\mathbf{s}_j(\mathcal{C})$ using the same $K_{\mathrm{seq}}$ and correlates it with the observed residual vector to recover the bit sign. Even if an adversary knows the overall embedding domain and how candidates are constructed, without $K_{\mathrm{seq}}$ it remains difficult to reconstruct the bit templates required for decoding or matched removal.

\subsection{Spread Spectrum Embedding Algorithm}
\label{subsec:multi-bit-payload-construction-redundancy-design}
In our baseline experimental setting, the spherical-harmonics parameters of 3DGS are selected as the embedding domain. Compared with other Gaussian parameters, center coordinates with dimension $3$, opacity with dimension $1$, rotation with dimension $4$, and scale with dimension $3$, the SH parameters provide a $48$-dimensional embedding space. Moreover, other Gaussian parameters are tightly coupled with scene geometry, which means that small perturbations can severely degrade the overall scene fidelity. Therefore, restricting modifications to the SH parameters is more favorable, and is also a common choice in the existing works.

For each scene, all SH parameters of all Gaussians are concatenated into a sequence of 1D length, and a 1D DCT is applied along the Gaussian index axis to obtain transform-domain coefficients. We then select a mid-band interval of DCT frequency indices as candidate embedding locations to form the transform-domain candidate carrier pool $\mathcal{P}$, and denote its size by $T = |\mathcal{P}|$. To make the embedding layout key-dependent and reproducible, $K_{\mathrm{sel}}$ is used to generate a keyed permutation $\pi$ over carrier indices.

Using $K_{\mathrm{code}}$, a bipolar payload vector $\mathbf{b}(\mathcal{C})\in\{-1,+1\}^{B}$ is derived. For each bit position $j\in\{1,\ldots,B\}$, a length-$T$ bipolar spreading template $\mathbf{s}_j(\mathcal{C})\in\{-1,+1\}^{T}$ is generated using $K_{\mathrm{seq}}$. Following that setting, we adopt a code-division superposition embedding in the carrier domain and define the overall perturbation vector as
\[
\boldsymbol{\delta}(\mathcal{C})=\sum_{j=1}^{B} \alpha_j\, b_j(\mathcal{C})\,\mathbf{s}_j(\mathcal{C}),
\]
where $\alpha_j$ denotes the per-bit embedding amplitude. Finally, after injecting the perturbation, the updated transform-domain representation is mapped back to the parameter domain, producing the watermarked model $\mathcal{M}_w$.

\subsection{3DGS Model Watermark Detector}
\label{subsec:fidelity-control-imperceptibility-analysis}
On the 3DGS-level detection side, we follow the classical non-blind detection protocol widely used in traditional media. Given access to the original model $\mathcal{M}$ and a suspect model $\widetilde{\mathcal{M}}$, we construct the transform-domain residual by taking DCT coefficient differences at carrier locations in $\mathcal{P}$. Using $K_{\mathrm{sel}}$ and the claim $\mathcal{C}$, the detector reproduces the same keyed permutation $\pi$. We further adopt a detector-side carrier budget by operating on a deterministic subset $\Omega \subseteq \{1,\ldots,T\}$ of carrier indices in order to reduce computational cost and standardize the detection overhead. Specifically, $\Omega$ is sampled from the permuted carrier order in a key-dependent and reproducible manner using $K_{\mathrm{sel}}$ and the claim $\mathcal{C}$.

For each bit position $j$, the detector generates the spreading template $\mathbf{s}_j(\mathcal{C})$ using $K_{\mathrm{seq}}$, extracts its subvector $\mathbf{s}_{j,\Omega}(\mathcal{C})$, and then computes the correlation score
\[
S_j=\langle \mathbf{z}_{\Omega},\,\mathbf{s}_{j,\Omega}(\mathcal{C})\rangle,
\]
where $\mathbf{z}_{\Omega}$ is the residual observation vector extracted according to the permuted carrier order and the budgeted subset. Finally, the bit can be recovered via a sign test
\[
\hat b_j=
\begin{cases}
+1, & S_j \ge 0,\\
-1, & S_j < 0.
\end{cases}
\]

%%%%%%%%%%%%%%%%%%%%%%%%%%%%%%%%%%%%%%%%%%%%%%%%%%%%%%%%%%%%%%%%%%%%%%%%%%%%%%%
%%%%%%%%%%%%%%%%%%%%%%%%%%%%%%%%%%%%%%%%%%%%%%%%%%%%%%%%%%%%%%%%%%%%%%%%%%%%%%%

\section{Analysis of the Existing Works on 3DGS watermarking}
\label{Appendix_C}

In Table~1, we decompose representative existing works and summarize them along several axes. Bit String indicates the information payload carried by the watermark (in bits). It directly governs two core trade-offs, payload--robustness and payload--fidelity. In this paper, we advocate reporting results under $32/48/64$-bit payloads. Too short payloads may only support coarse identifiers and can be insufficient for richer claims, making them less informative in practice~\cite{chen2025guardsplat}. Moreover, reporting a single payload length can mislead subsequent readers about the flexibility of payload configuration and obscure the capacity-related trade-offs~\cite{li2025gs,tan2024water,huang2024gaussianmarker}.

For the Embedding Domain, gradient-based optimization in neural rendering can preserve the fidelity of the watermarked model, which motivates many methods to embed watermarks globally across the representation~\cite{jang20253d,li2025gs,tan2024water,huang2024gaussianmarker}. However, it is crucial to note that when watermark embedding is performed with some computing methods, certain Gaussian parameters, such as positions and orientations, are coupled with scene geometry. Even small perturbations to these geometry-sensitive parameters can cause disproportionate geometric artifacts and visible distortions.

For the Embedding Stage, post-hoc optimization is among the most widely adopted and stable paradigms, including in NeRF-based watermarking. Nevertheless, it typically incurs substantial optimization time. In contrast, single-forward schemes avoid iterative optimization, but often rely on cross-domain transformations, such as from 3DGS parameters to intermediate images representations, which can introduce information loss.~\cite{huang2025marksplatter}. Existing 3DGS-level single-forward schemes still remain less mature in robustness and security, leaving substantial room for improvement~\cite{li2025gs}.

The detector design for rendered images raises a fundamental security concern. Several works adopt HiDDeN~\cite{zhu2018hidden} as the detector for 2D rendered watermarks~\cite{jang20253d,li2025gs,tan2024water,huang2024gaussianmarker}, implying that detection is publicly runnable. Meanwhile, none of these works explicitly incorporate constraints such as secret keys or private parameters, which exposes them to standard security risks such as reverse engineering, adaptive querying, and forgery. In particular, if the detector is public, an attacker can replicate the pipeline and mount gradient-based attacks to selectively erase or counterfeit the watermark.

The datasets used in prior studies can be broadly categorized into three groups: synthetic scenes, real-world scenes, and large-scale object-centric data. Synthetic scenes like Blender~\cite{mildenhall2021nerf} offer controlled factors and aligned quantitative evaluation, but can exhibit distributional bias and under-represent real capture artifacts. Real-world scenes contain LLFF~\cite{mildenhall2019local}, Mip-NeRF~360~\cite{barron2022mip}, and Tanks\&Temples~\cite{knapitsch2017tanks}, which better reflect realistic capture noise and complex appearance, but make training and rendering settings harder to standardize across methods. Large-scale object-centric datasets have Objaverse~\cite{deitke2023objaverse}, OmniObject3D~\cite{wu2023omniobject3d}, and Google Scanned Objects~\cite{downs2022google}. They emphasize asset diversity and large-scale generalization, which are suitable for stress-testing stability across many objects and categories. However, most samples depict isolated objects with simple or missing backgrounds, lacking global illumination, occlusions, complex backgrounds, and inter-object light transport effects commonly present in real scenes.

In Table~2, the threat models simulated by all the aforementioned works are listed. We use upright font for attacks applied to rendered 2D media, such as Gaussian Noise ($\sigma=0.1$), and italic font for attacks applied directly to the 3DGS model, such as \textit{Gaussian noise ($\sigma=0.1$)}. For each attack, the attacking domain and the corresponding evaluation metrics are also specified.

% We reiterate the following statement here, ``No single watermarking framework can be perfect for all scenarios. Without scenario-grounded definitions, watermarking applications can easily collapse into engineering-oriented data hiding and fail to address the core security questions.'' These existing threat models can continue to provide engineering guidance for subsequent studies. However, it should be clear that merely increasing the number of threat models does not necessarily make a proposed watermarking framework more secure. Practical effectiveness is achieved only after the application scenario is clearly defined, and the threat model is then designed to match the adversarial capabilities and objectives implied by that scenario.

\section{Trade-off Between Fidelity and Different Payload Lengths}
\begin{figure*}[t]
  \centering
  \includegraphics[width=\textwidth]{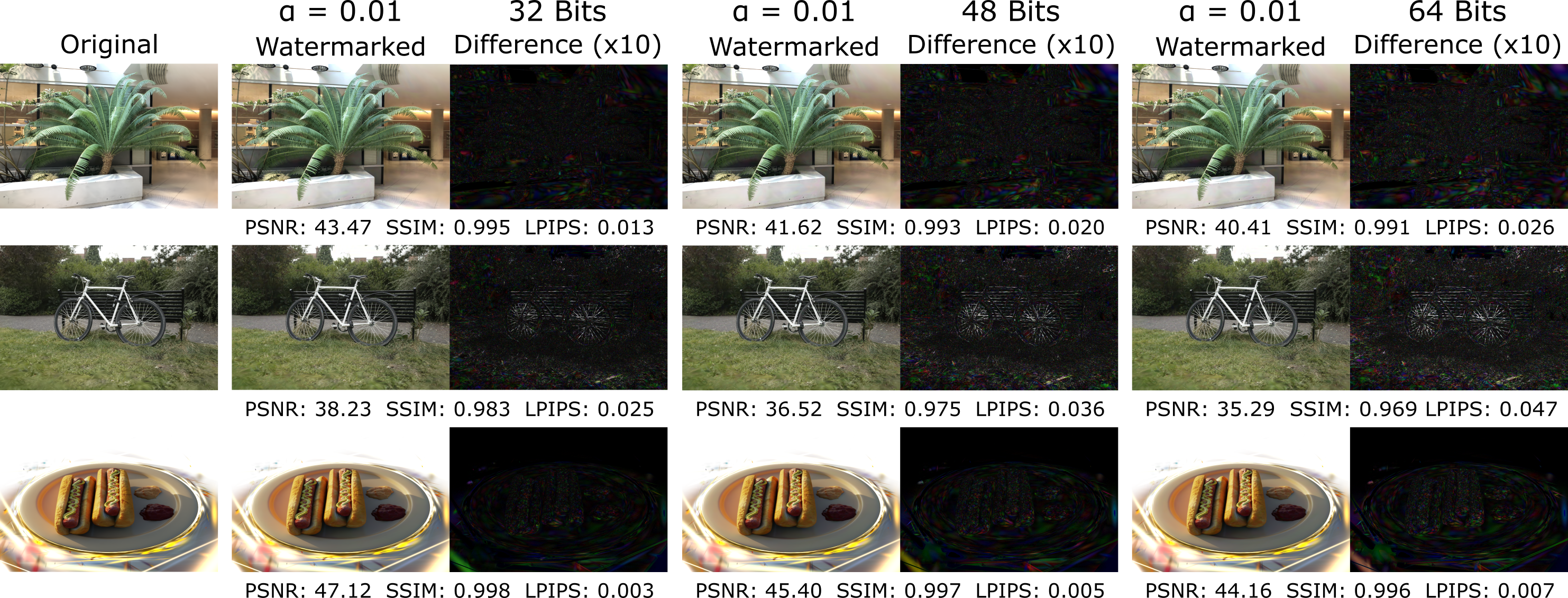}
  \caption{Qualitative fidelity with different payload lengths. Rendered original, watermarked, and difference ($\times 10$) images are represented with a fixed embedding strength $\alpha=0.01$ and payload lengths $B \in \{32,\,48,\,64\}$. The three rows correspond to LLFF \emph{fern}, Mip-NeRF~360 \emph{bicycle}, and Blender \emph{hotdog}, respectively.
}
  \label{Appendix_D}
\end{figure*}

Figure~4 illustrates the trade-off between the robustness and payload under a fixed total embedding energy constraint using additive Gaussian noise with $\sigma=0.3$. Detection accuracy decreases as the payload increases, since a fixed total embedding energy budget allocates less energy to each bit, which in turn weakens robustness under noise.

\end{document}